\documentclass{article}

\usepackage[top=1in, bottom=1in, left=1in, right=1in]{geometry}

\usepackage{microtype}
\usepackage{graphicx}
\usepackage{parskip}
\usepackage[numbers]{natbib}
\usepackage{booktabs}

\usepackage{hyperref}
\usepackage{color}
\usepackage{wrapfig}
\usepackage{multirow}

\newcommand{\lb}[2][l]{%
  \begin{tabular}[#1]{@{}l@{}}#2\end{tabular}}

\usepackage{amsmath}
\usepackage{amssymb}
\usepackage{algorithm}
\usepackage{algorithmicx} 
\usepackage[noend]{algpseudocode} 
\usepackage{graphicx}
\usepackage{amsthm}
\usepackage{float}
\usepackage{physics}
\usepackage{subfig}
\usepackage{xcolor}

\newcommand{\E}{\mathbb{E}}

\newcommand{\hy}{\hat{y}}
\newcommand{\der}{\delta_{reward}}
\newcommand{\dep}{\delta_{penalty}}
\newcommand{\dc}{\text{disc}}

\newcommand\R{\mathbb R}

\newtheorem{theorem}{Theorem}[section]

\newtheorem{claim}[theorem]{Claim}

\def\h_#1{\hat{#1}}
\def\wh_#1{\widehat{#1}}

\newcommand\set[1]{\left\{#1\right\}} %

\usepackage[textsize=scriptsize,backgroundcolor=green]{todonotes}

\newenvironment{varalgorithm}[1]
  {\algorithm\renewcommand{\thealgorithm}{#1}}
  {\endalgorithm}

\newfloat{algorithm}{H}{}

\usepackage{cleveref}
\crefformat{footnote}{#2\footnotemark[#1]#3}

\title{\bf Designing Closed-Loop Models for Task Allocation}
\author{Vijay Keswani \\ Yale University \and L. Elisa Celis \\ Yale University \and Krishnaram Kenthapadi \\ Fiddler AI \and Matthew Lease \\ University of Texas at Austin}
\date{}

\begin{document}
\maketitle

\begin{abstract}
Automatically assigning tasks to people is challenging because human performance can vary across tasks for many reasons. This challenge is further compounded in real-life settings in which no oracle exists to assess the quality of human decisions and task assignments made. Instead, we find ourselves in a ``closed'' decision-making loop in which the same fallible human decisions we rely on in practice must also be used to guide task allocation. How can imperfect and potentially biased human decisions train an accurate allocation model? Our key insight is to exploit weak prior information on human-task similarity to bootstrap model training. We show that the use of such a weak prior can improve task allocation accuracy, even when human decision-makers are fallible and biased. 
We present both theoretical analysis and empirical evaluation over  synthetic data and a social media toxicity detection task. Results demonstrate the efficacy of our approach.
\end{abstract}

\section{Introduction} \label{sec:introduction}

Human decision-making is ubiquitous: in the daily life of organizations or ``pure'' {\em human computation} settings without automation, in making labeling decisions to train and test AI systems, and in {\em human-in-the-loop} architectures that dovetail automated AI with human abilities.
People are also naturally fallible:
some people perform better than others across different tasks due to a wide range of factors (e.g., background or experience), as observed in recruitment \cite{bertrand2004emily} and healthcare \cite{raghu2019direct}. Human error can be due to noise (e.g., fatigue/oversight) and systematic patterns of error (e.g., varying skill). %
Group decisions can also be fallible and systematically biased  depending on the composition and decision process. Whereas ``wisdom of crowds'' \cite{surowiecki2005wisdom} can boost collective intelligence via group diversity, lack of such diversity can amplify biases rather than mitigate them \cite{gorwa2020algorithmic}.

\begin{figure*}[t]
    \centering
    \includegraphics[width=\linewidth]{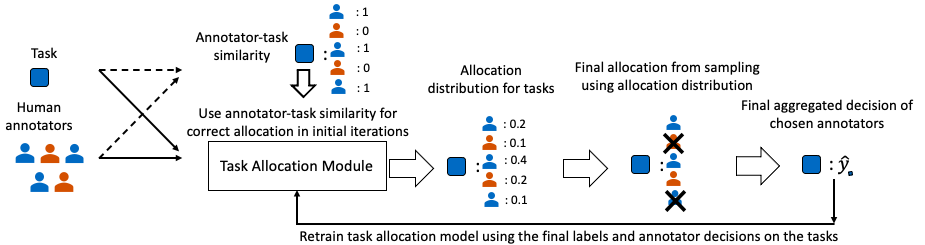}
    \caption{A closed-loop task allocation model in which predictions are fed back to train the model. To bootstrap training, we use prior information on human-task similarity, evaluated here by matching task \& annotator colors.}
    \label{fig:process}
\end{figure*}

{\em Task allocation} (cf.\ \cite{hettiachchi2020crowdcog}) seeks to optimize the overall quality of outcomes by effectively matching people to tasks.
Accurate task allocation has applications in crowdsourcing \cite{fan2015icrowd}, human-in-the-loop frameworks \cite{keswani2021towards}, and collaborative web platforms \cite{anagnostopoulos2012online}.
A key assumption underlying most prior work on task allocation is that an oracle exists to provide feedback on the quality of human decisions and task assignments made. In real life, however, the same fallible human decisions we rely on must often also provide the basis for evaluating  allocation decisions.
When a hiring or admissions committee makes a decision whether to hire/admit a given candidate, all we have is the committee's decision; no 
outside oracle exists to provide a definitive evaluation of the committee's decision. 
Similarly, social media content moderation relies on decisions from human moderators.
Moreover, decision criteria are often organization-specific
model \cite{pan2021adoption}. 
These applications motivate our investigation of the \textit{closed} training loop setting in which the aggregated annotations from an input-specific selection of human decision-makers are fed back into the system to train the task allocation model (\textbf{Figure~\ref{fig:process}}).  
However, considering that human decision-makers can make imperfect decisions, the question arises whether their aggregated decisions can be used to train an accurate task allocation framework.

A central challenge with such a framework is how to address human inaccuracy and bias, especially in the initial training iterations. Unsupervised aggregation of human decisions \cite{dawid1979maximum} can provide noisy feedback on task allocation efficacy \cite{goel2019crowdsourcing,fan2015icrowd,liu2012cdas,wu2021task,tran2014efficient}; however %
such noise, especially in initial training iterations, can result in a slow or non-converging training process. Furthermore, any bias in human decisions may be fed back into task allocation training, further amplifying system error \cite{mehrabi2019survey} (see \S\ref{sec:training_overview}). Particularly problematic are human biases stemming from a lack of background, training, or prejudice, which can consistently impair %
performance. %
Another factor that can influence human decisions is underlying demographic identity. 
\citet{goyal2022your} observe that the demographic identity of crowd annotators impacts their toxicity ratings.
Consequently, they call for ``\textit{ML engineers and researchers to ... consider all the different kinds of specialized rater pools we should be imagining, considering, designing, and testing with.}'' 
Multiple other studies \cite{kumar2021designing, bhuiyan2020investigating, gordon2022jury, spinde2021mbic,sap2021annotators} have reported significant differences in ratings across annotator demographics (see \S\ref{sec:related_work} for additional related work). 
Motivated by these studies, we tackle the problem of developing allocation methods that are input-specific, contextually-aware, and cognizant of the background of the human annotators.

\noindent
\textbf{Contributions.} In this work, we formulate the challenge of closed-loop learning from noisy human labels and propose two online learning algorithms for it. To mitigate inaccuracy from fallible human decision-makers, we propose exploiting a specific form of weak prior on human-task similarity to initialize the task allocation model (\S\ref{sec:prior}). This enables us to obtain relatively accurate class labels for initial inputs and thereby effectively bootstrap the training process. The first algorithm we present, {\em \ref{alg:main}}, directly uses the prior information to initialize the allocation model. The second, {\em \ref{alg:main_2}}, provides a smoother transition from the prior distribution to learning from noisy feedback during training. 
We demonstrate the efficacy of our methods via both theoretical analysis (\S\ref{sec:algorithms}) and empirical improvement on synthetic and real-world datasets (\S\ref{eval-syn-ta} and \S\ref{eval-real}). The latter extends beyond the classic assumption of universal, objective truth to consider recent advocacy for recognizing subjective, community-based gold standards
\cite{sen2015turkers,kumar2021designing,gordon2022jury,goyal2022your}.

\section{Model and Algorithm} \label{sec:model}

We consider the binary classification task of predicting label $y$ from input $x \in \R^n$.
We assume each input $x$ belongs to one or more categories $z \in \mathcal{Z}$, which could correspond to any demographic or task-specific interpretable feature.
Given any input $x$ (i.e., a task), the goal of task allocation is to choose appropriate human annotators (interchangeably referred to as individuals or decision-makers) from a given pool, whose aggregated prediction is the final predicted label for $x$.
Assume there is a pool of $m$ available annotators $e_1, \dots, e_m : \R^n \rightarrow \set{0,1}$, with $e_i(x)$ denoting the $i$-th annotator's prediction for $x$.
For input $x$, the task allocation model $D_u$ infers a probability distribution over annotators: $D_u: \R^n \rightarrow \Delta^m$, where $u$ denotes model parameters and $\Delta^m$ denotes the $m$-dimensional simplex\footnote{distribution over $m$ annotators: $\forall d{\in}\Delta^m$, $0{\leq}d_i{\leq}1$ for all $i{\in}\set{1, \dots, m}$ and $d^\top \mathbf{1}{=}1$.}. When possible, we omit subscript $u$ and refer to $D_u$ by $D$. $D(x)_i$ denotes the model probability assigned to individual $i$, reflecting the model's estimate of that individual's ability to correctly label input $x$, relative to the other annotators.
While not evaluated in our study, our framework also supports each person having additional input-specific costs associated with their predictions
(see discussion of this point in \S\ref{sec:discussion}).

\noindent
\textbf{Committee Voting.} Given the task allocation model $D(x)$'s inferred probability distribution over the pool of $m$ annotators, the top-$k$ can be selected to form a committee. When $k>1$, the committee's decision is determined by majority vote (assuming $k$ is odd, no tie-breaking is required). A technical detail is that we sample annotators with replacement according to $D(x)$ so that the the majority vote of the committee implicitly equals (on average) the weighted majority vote of all $m$ annotators, with $D(x)$ probabilities as weights. Alternatively, one could sample the $k$ annotators without replacement and explicitly weight member votes by $D(x)_i$.

\noindent\textbf{Online learning.} Assuming a streaming setting, after each input $x$ is labeled by a selected committee, the (potentially noisy) label is fed back into the closed-loop learning process to update the model $D(x)$. This online learning setting supports potential use in various real-world applications \cite{fontenla2013online, awerbuch2008online}. However, such a noisy feedback loop also risks problematic predictions when trained without care;
our algorithms are thus designed to address this.

\subsection{Training the Allocation Framework} \label{sec:training_overview}

An ideal training process for an allocation framework learns a partition of the feature space and assigns annotators to those partitions where they are expected to be most accurate.
Prior training approaches optimize over labeled datasets to learn an allocation model that simulates such a partition \cite{keswani2021towards,tran2014efficient,goel2019crowdsourcing}.
In this section, we first summarize training procedures from prior work (that assume access to oracle training labels or rewards/penalties). We then discuss extensions of these procedures for closed-loop training.

\setlength{\abovedisplayskip}{3pt}
\setlength{\belowdisplayskip}{3pt}

\noindent
\textbf{Prior work training allocation models with gold.}
Assume input $x$ having group attribute $z$ and true binary label $y$, $D(x)$ is the task allocation model probability distribution over the pool $m$ experts, and ${e_i}(x)$  binary prediction of expert $i$. A general training algorithm, with access to ground truth labels, will update the allocation model to reward the correct experts (for whom $e_i(x) = y$) and penalize those who are incorrect:
\begin{equation}
D(x)_i = \begin{cases} D(x)_i + \delta_{reward}^{(i)}(x,y),  &\text{ if } {e_i}(x) = y\\ D(x)_i - \delta_{penalty}^{(i)}(x,y),  &\text{ if } {e_i}(x) \neq y \end{cases} 
\label{updates-True} 
\end{equation}
where $\delta_{reward}^{(i)}(\cdot), \delta_{penalty}^{(i)}(\cdot) : \mathcal{X} \times \set{0,1} \rightarrow \R_{\geq 0}$ are input and annotator-specific updates and chosen so that the updated weights sum to 1\footnote{i.e., $\sum_{i=1}^m \delta_{reward}^{(i)}(x,y) \cdot \mathbf{1}(e_i=y) - \sum_{i=1}^m \delta_{penalty}^{(i)}(x,y) \cdot \mathbf{1}(e_i\neq y) = 0$.}.
This appropriately rewards/penalizes the annotators, yielding allocation model updates that simulate these rewards/penalties.

The reward/penalty functions are constructed by framing the problem as an optimization program. 
In the case of \citet{keswani2021towards}, rewards/penalties are constructed as follows: given $(x,y)$, allocation parameters $u$, and committee size $k$, first select a committee $C$ of $k$ annotators using $D_u(x)$ and compute the (probabilistic) prediction $\hat{y}_{u}(x)$ by taking the mean of the selected annotators, i.e., $\hat{y}_{u}(x) := \sum_{i \in C} e_i(x) / |C|$.
Then minimize the following regularized log-loss function: 
$\mathcal{L}_D(u) :=\E_{x,y}\left[ -y \log(\sigma(\hat{y}_{u}(x))) - (1-y)\log (1 - \sigma(\hat{y}_{u}(x)))) \right],$ where $\sigma$ is the standard sigmoid function.
Expected loss can be computed by the mean over a batch of training samples, with optimization performed via gradient descent.
The gradient updates for this loss function can be seen to reward the correct annotators and penalize the incorrect annotators \cite{keswani2021towards}. Hence, functionally, each step of this algorithm has a similar structure as {\bf Equation~\ref{updates-True}}. 
Other prior training algorithms can also be shown to have similar underlying reward/penalty structure;
see Appendix~\ref{sec:prior_training} for examples.

\noindent
\textbf{Training using noisy aggregated human labels.} In this work, we focus on the more challenging case of having access to fallible human decisions only, with no oracle feedback regarding their accuracy (i.e., no access to $y$).
Lacking gold labels, one way to directly use the above training process is to learn from noisy, aggregate human labels.
Given input $x$ and committee $C$ selected using $D(x)$, the predicted label $\hy(x) := \mathbf{1}\left[\sum_{i \in C} e_i(x)/|C| > 0.5\right]$.
Then, the training updates can substitute $y$ with $\hat{y}$ in Equation~\ref{updates-True}:
\begin{equation}
D(x)_i = 
\begin{cases} D(x)_i + \delta_{reward}^{(i)}(x,\hy(x)),  &\text{ if } {e_i}(x) = \hy(x)\\ D(x)_i - \delta_{penalty}^{(i)}(x,\hy(x)),  &\text{ if } {e_i}(x) \neq \hy(x) 
\end{cases} 
\label{updates-Noisy}
\end{equation}
By substituting true class labels with noisy aggregated labels,   existing training allocation algorithms \cite{keswani2021towards, mozannar2020consistent} can be used without major changes (e.g., substitute $y$ with $\hat{y}$ in above loss $\mathcal{L}_D(u)$). 
While simple, this approach also has a potential downside: 
when the majority of the annotators are consistently biased against any group  $z\in\mathcal{Z}$, this unsupervised training process is unable to detect such bias.

\noindent
\textbf{Bias propagation when training using noisy labels.}
Assuming a binary group attribute, we show below that: if (i)  the starting allocation model chooses annotators randomly, and (ii) the majority of the annotators are biased against or highly inaccurate with respect to a group attribute type (e.g., a disadvantaged group), then 
the above training process leads to disparate performance with respect to the disadvantaged group.
For $\alpha> 0.5$, assume that $\alpha$ fraction of annotators are biased against group $z=0$ and $(1{-}\alpha)$ fraction are biased against group $z=1$.
If a person 
is biased against $z=j$, they will always predict correctly for inputs with $z{=}1{-}j$ but predict correctly for inputs with $z{=}j$ with probability 0.5.

Lacking an informative prior, training will start with $D(x)$ assigning uniform probability $1/m$ to all $m$ annotators. 
When $k=1$, a single person decides the label for input $x$.
In this case, %
the starting accuracy for group $z=1$ elements will be $\alpha + 0.5(1-\alpha)$, and for group $z=0$ elements, $(1-\alpha) + 0.5\alpha $. Therefore, the difference in expected accuracy for group $z=1$ vs.\ $z=0$ elements will be $(\alpha - 0.5)$. The larger the value of $\alpha$, the greater the disparity will be.
Hence, with biased starting allocation model and predicted labels used for retraining, the bias will propagate to the learned model. %

\begin{claim} \label{clm:example}
In the above setting, the disparity between accuracy for group $z{=}0$ and accuracy for group $z{=}1$ does not decrease even after training using multiple Eqn.~\ref{updates-Noisy} steps.
\end{claim}

\noindent
The proof is provided in Appendix~\ref{sec:proofs}. In \S\ref{eval-syn-ta}, we simulate a setting wherein most annotators are biased against certain input categories. Results show that prior training algorithms perform poorly, yielding low allocation accuracy.

\subsection{Injecting Prior Information} \label{sec:prior}

In real-life, no oracle exists to provide us feedback on our fallible or biased human decisions. There is no oracle gold training data to guide initial allocation decisions, nor is there gold feedback on human decisions made during closed-loop training.  
How then can imperfect human decisions train an accurate task allocation model? Our key insight is to exploit weak prior information on human-task similarity to bootstrap model training.

\noindent
\textbf{Motivating Examples.}
\textit{Example 1.} When a company recruits a new employee, the human decision-makers are typically current employees, and more specifically, a ``hiring loop'' of employees possessing appropriate expertise to assess the candidate's credentials. Assuming a company knows the varying expertise of its own workforce, prior information exists to match decision-makers to new candidates.
In addition, organizations today appreciate the importance of forming hiring committees that combine  diversity and expertise \cite{schumann2020we}. 

\textit{Example 2.} In content moderation, moderator decisions vary due to many compatibility factors.
For example, lack of familiarity with the dialect of the content's author can lead to biased decisions \cite{sap2019risk,davidson2019racial}. Whether the moderator has themself been a target of hate speech \cite{kumar2021designing}, or whether their own demographic identity aligns with that being targeted in content they are reviewing \cite{goyal2022your} can also impact their decisions. Thus, once differences in judging behavior among moderators is acknowledged and accepted, it creates a space for matching different groups of moderators to different content types, based on moderator background (which can be collected via an on-boarding questionnaire).

\noindent
\textbf{Encoding Prior Information.}
Any allocation model induces a probability distribution over the decision-makers for each input, such that the probability assigned to each decision-maker represents the confidence in their correctness. %
An initial approximation of this distribution over the human decision-makers can be derived using the contextual information of the application where the allocation model is being employed.
For the motivating examples above, such weak prior information already exists to 1) appoint employees to a hiring loop who are capable of evaluating a candidate (by matching areas of expertise); and 2) select moderators to review content appropriate to their background (by matching target and annotator demographics/dialect).

In absence of labeled training data, we can use this prior information to bootstrap the closed-loop training process.
The prior information is encoded in our framework using a similarity function $dSim : \set{e_1, \dots, e_m} \times \mathcal{Z} \rightarrow [0,1]$, 
i.e., specifying a continuous similarity score
matching each individual person to each content category. 
As shown above, starting with a random allocation model is challenging when we also lack oracle feedback on the accuracy of human decisions in the closed-loop training process. 
Especially problematic are settings when the majority of annotators are biased against certain groups, as observed from the stylized example in Claim~\ref{clm:example}.
By starting with some prior information about which people (or groups of people) might be best suited to each type of task (or category of content) using $dSim$, we seek to address this flaw of the closed training framework and bootstrap an accurate training process.
Indeed Claim~\ref{rem:example_dsim}, shows that using an appropriate $dSim$ can address the issues observed in the setting of Claim~\ref{clm:example}.

\begin{claim} \label{rem:example_dsim}
Revisiting 
Claim~\ref{clm:example}, suppose %
$dSim(e_j, z){=}1$ if annotator $e_j$ is unbiased for category $z$ and $\gamma$ otherwise, where $\gamma$ is any constant $\in [0,1]$.
Consider the allocation induced by this $dSim$ function (i.e., for input $(x,z)$, allocation output $D(x)_i \propto dSim(e_i, z)$). 
Then the difference between the accuracy for group $z{=}0$ and $z{=}1$ lies in $\left[ \frac{\gamma}{2}, \frac{\alpha}{1-\alpha} \cdot \frac{\gamma}{2} \right]$.
\end{claim}

\noindent
The proof is provided in Appendix~\ref{sec:proofs}. Claim~\ref{rem:example_dsim} shows that smaller $\gamma$ values imply $dSim$ is better able to differentiate between biased and unbiased annotators.
Hence, %
the better $dSim$ is at differentiating biased and unbiased annotators, the smaller is the disparity in performance across groups of the starting allocation model. 
We thus utilize $dSim$ to mitigate biases in training using noisy labels (i.e., \textbf{Eq.~\eqref{updates-Noisy}}). 
Our proposed algorithms in the next section operate on this general formulation of $dSim(e,z)$. 

\begin{algorithm}[H]
  \caption{Training with prior information.\\
  {\bf Input}: inputs $(x_1,z_1), \dots (x_T, z_T)$,  annotators $e_1, \dots, e_{m}$, parameter $k$ \\ {\bf Output:} Trained task allocation model}
  \begin{algorithmic}[1] 
	\State Set initial allocation $D_{0}$ such that for any input $x$ in category $z$, we have $D_{u_0}(x)_i \propto dSim(e_i, z) $
  	\For{$t \in \{1, 2, \dots,  T\}$}
        \State $D_{{t-1}}(x_t) \gets $ Allocation distribution for  $x_t$
        \State $C \gets$ Choose committee of size $k$ using distribution $D_{{t-1}}(x_t)$
    	\State $\hat{y}_t \gets $ Aggregated predictions of annotators in committee $C$
		\State $D_t \gets$ Update allocation by training on $(x_t, \hat{y}_t)$
    \EndFor    
	\State return $D_T$
\end{algorithmic}
    \label{alg:main_overview}
\end{algorithm}

\subsection{Training a Closed-loop Framework using $dSim$} \label{sec:algorithms}

Algorithm~\ref{alg:main_overview} presents our general training process.
The first step ensures that initial allocation follows the prior information provided by $dSim$.
Subsequent training steps learn from noisy, aggregated decisions to further improve the task allocation model accuracy.
Concrete methods to implement this algorithm are discussed next.

\noindent
\textbf{Training Method 1: \ref{alg:main}.} \label{sec:preproc_algorithm}
One way to implement Algorithm~\ref{alg:main_overview} in practice is to encode the $dSim$ function within the initial task allocation model.
\begin{varalgorithm}{Strict-Matching}
\flushleft
  \caption{\\{\bf Input}: input stream $(x_1,z_1),\dots (x_T,z_T)$,  experts $e_1, \dots, e_{m}$, function $dSim$, batch size $B$, committee size $k$, parameter $T_d$, rate $\eta$, loss function $\mathcal{L}_D$.\\ {\bf Output:} Trained allocation model parameters $u$.}
  \begin{algorithmic}[1] 
	\State Set initial model parameters $u_0$ s.t. for any input $(x,z)$, we have $D_{u_0}(x)_i \propto dSim(e_i, z) $
	\State $S \gets \emptyset$
  	\For{$t \in \{1, 2, \dots,  T\}$}
        \State $D_{u_{t-1}}(x_t) \gets $ Allocation output for $x_t$
        \State $D(x_t) \gets D_{u_{t-1}}(x_t)/$sum$(D_{u_{t-1}}(x_t))$ 
    	\State $C \gets$ Sample $k$ annotators from $D(x_t)$
    	\State $\hat{y}_t \gets $ Aggregate label of committee $C$  
    	\State $S \gets S \cup \set{(x_t, y_t)}$
    	\If{$|S| = B$}
            \State $u \gets u - \eta \cdot \pdv{\mathcal{L}_D(u)}{u}\Big{|}_{S}$     
		    \State $S \gets \emptyset$
		\EndIf
	\EndFor
	\State return $u_T$
\end{algorithmic}
    \label{alg:main}
\end{varalgorithm}
In particular, we set initial allocation model parameters such that, for the starting allocation model $D_{u_0}$ and input $(x,z)$ and annotator $e_i$, we have that $D_{u_0}(x)_i \propto dSim(e_i, z) $ (Step 1).
This can be feasibly accomplished in most applications using unlabeled data.
The rest of the training process is the same as \S\ref{sec:training_overview} and Equation~\eqref{updates-Noisy}: for every input, reward the annotators whose prediction matches with aggregated prediction and penalize those who do not (using gradient of loss $\mathcal{L}_D$). %
Aggregation of selected annotator predictions can be implemented in various ways; see {\em Committee Voting} in \S\ref{sec:model}.
To add further robustness, we use a batch update process; i.e., for a given integer $B$, train the model after observing $B$ samples. 
This approach exploits the $dSim$ prior to set the initial $D(x)$ distribution, %
followed by closed-loop training with noisy aggregate feedback to further improve the task allocation model.
For a given input (category), if an annotator shows low $dSim$ score  but high observed accuracy (or vice-versa), the allocation model can learn this property as training progresses. 
Consequently, we can view this training regime as a simple form of exploration-exploitation.

\noindent
\textbf{Training Method 2: Smooth-Matching.} \label{sec:ee_algorithm}
To obtain a better transition from the $dSim$ prior to the allocation model learnt during closed-loop training, we can gradually wean ourselves off of the prior by decreasing its relative weight as more observed evidence accumulates.
\begin{varalgorithm}{Smooth-Matching}
\flushleft
  \caption{\\{\bf Input}: input stream $(x_1,z_1),\dots (x_T,z_T)$,  experts $e_1, \dots, e_{m}$, function $dSim$, batch size $B$, committee size $k$, parameter $T_d$, rate $\eta$, loss function $\mathcal{L}_D$.\\ {\bf Output:} Trained allocation model parameters $u$.}
  \begin{algorithmic}[1] 
    \State $S \gets \emptyset$
  	\For{$t \in \{1, 2, \dots,  T\}$}
	        \State $\mu \gets {T_{d}}/(t+T_{d})$
	        \State {\footnotesize $D_{dSim}(x_t) \gets [dSim(e_1, z_t), ... , dSim(e_{m},z_t)]$}
	        \State {$D_{dSim}(x_t) \gets D_{dSim}(x_t)/\text{sum}(D_{dSim}(x_t))$}
            \State $D_{u_{t-1}}(x_t) \gets $ Allocation output for $x_t$
            \State $D(x_t) \gets D_{u_{t-1}}(x_t)/$sum$(D_{u_{t-1}}(x_t))$ 
	        \State {$D_{comb} \gets \mu \cdot D_{dSim}(x_t) + (1-\mu) \cdot D(x_t)$}	        
        	\State $C \gets$ Sample $k$ experts from $D_{comb}$
        	\State $\hat{y}_t \gets $ Aggregate label from committee $C$  
        	\State $S \gets S \cup \set{(x_t, y_t)}$
        	\If{$|S| = B$}
                \State $u \gets u - \eta \cdot \pdv{\mathcal{L}_D(u)}{u}\Big{|}_{S}$ 
    		    \State $S \gets \emptyset$
    		\EndIf
    	\EndFor    
	\State return $u_T$
\end{algorithmic}
    \label{alg:main_2}
\end{varalgorithm}
In other words, the allocation employed at any iteration can be chosen as a convex combination of the allocation encoded by the $dSim$ prior and the allocation trained using the observed samples (and aggregated class labels).
This method of combining prior and observed data is conceptually similar to Bayesian or Laplacian smoothing techniques \cite{schutze2008introduction,valcarce2016additive}. Additive combination yields a task allocation distribution incorporating both the prior distribution and the empirical distribution.
The smoothing parameter $\mu$ is set to be an increasing function of the number of observations, ensuring that prior information $dSim$ is used primarily in the initial training iterations.
Full details are provided in Algorithm~{\em\ref{alg:main_2}}.
Parameter $T_d$ in \ref{alg:main_2} controls the influence of $dSim$ on the training process.
The first $T_{d}$ iterations focus on obtaining accurate labels for initial samples to bootstrap the training process.
After $T_{d}$ iterations, the weight given to the prior is relatively smaller than the weight given to the distribution learned during training.

\subsection{Theoretical Analysis} \label{sec:theory}
Analyzing the two algorithms that use $dSim$ shows that the final trained allocation model simulates the underlying accuracy functions of the annotators.
The first theorem (\ref{thm:exploitation}) shows that if any annotator $e_j$ has high accuracy for category $z$, then how fast our algorithms converge to an allocation model that assigns high weight to $e_j$ for category $z$ depends on the initial weight assigned to $e_j$ for $z$.

\begin{theorem}[Exploitation using $dSim$] \label{thm:exploitation}
For any input group $z$, assume annotator $e_j$ is more accurate than all others. %
For $\beta > 0$, suppose we set $dSim$ function {\em st.} $dSim(e_j, z) - \max_{j' \in \set{1, \dots, m} \setminus \set{j}} dSim(e_{j'}, z) \geq \beta$.
Assume all annotators receive the same rewards/penalties for correct/incorrect predictions. 
Then the training algorithm that initializes $D(x)$ parameters $u$ with this $dSim$ function increases the weight assigned to annotator $e_j$ by at least $2\beta \delta$ in expectation, where $\delta \in [0,1]$ depends on the choice of $\der$ and $\dep$ values for the given input. 
\end{theorem}

\noindent
Hence, larger the $dSim$ weight for $e_j$, larger is their weight in the final allocation model.
Secondly, we show that when using appropriate $dSim$, if there are accurate annotators who are not assigned a high weight by $dSim$, they will be ``discovered'' during the training.

\begin{theorem}[Exploration of accurate annotators] \label{thm:exploration}
For any input group $z$, assume annotator $e_j$ has perfect accuracy ($1$).
Let $k$ be the size of the committee sampled from $D(x)$ to label input $x$.
Let the $dSim$ function be set such that $dSim(e_j, z) = \epsilon$, for some  $\epsilon \in [0,1]$, but the total weight (normalized) assigned by $dSim$ to accurate annotators for group $z$ is greater than 0.5. 
Assume all annotators receive the same rewards for correct prediction and same penalties for incorrect prediction. 
Then, there is an expected positive increase in the weight of this annotator if
$\epsilon > 1 - \left(1 - \frac{k}{2m} \right)^{1/k}$.
\end{theorem}
\noindent
Hence, our algorithms can discover accurate annotators so long as other accurate annotators are available to infer the true labels for this input category and $k, \epsilon$ are sufficiently large.
The proofs for both theorems are provided in Appendix~\ref{sec:proofs}.

\section{Evaluating Task Allocation on a Synthetic Dataset}  \label{eval-syn-ta}

Consider a binary classification task with three annotators having distinct areas of expertise, denoted by the colors {\color{orange}orange}, {\color{blue}blue}, and {\color{teal}green}. 
Assume each annotator is a perfect oracle when asked to label an example in their respective area but only 20\% accurate in the other two areas,
exhibiting consistent bias outside their respective areas of expertise. In the best case, the task allocator will correctly assign each input to the correct expert, yielding perfect labeling accuracy.  In the worst case, assigning every input to the wrong annotator will yield around 20\% accuracy. Because experts are assumed to be perfect oracles, each correct task allocation ensures a correct label. Consequently, task allocation accuracy largely determines label accuracy, which is lower-bounded by allocation accuracy.

\begin{figure*}
\begin{minipage}{\textwidth}
  \begin{minipage}[b]{0.4\textwidth}
    \centering
    \includegraphics[scale=0.45]{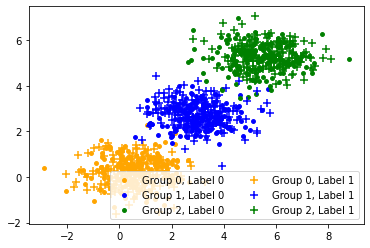}
    \captionof{figure}{Synthetic clusters in \S\ref{eval-syn-ta}.}
    \label{fig:syn_examples}
  \end{minipage}
  \hfill \quad
  \begin{minipage}[b]{0.59\textwidth}
  \begin{center}
    \begin{tabular}{lcc}
        \toprule
        Method & \lb{Label Acc.} & \lb{Assignment Acc.} \\
        \midrule        
        \ref{alg:main_2} & .90 (.08) & .87 (0.27) \\
        \ref{alg:main} & .79 (.01) & .74 (0.36)\\
        \midrule
        \citet{goel2019crowdsourcing} & .50 (.01) & .33 (.00) \\
        \citet{tran2014efficient} & .50 (.01) & .33 (.01) \\
        \citet{keswani2021towards} & .41 (.09) & .17 (.26) \\
        \bottomrule
    \end{tabular}
    \captionof{table}{Label and allocation accuracy for \S\ref{eval-syn-ta} with $s{=}0.3$. We report mean accuracy over 50 trials (standard error in brackets).} %
    \label{tbl:syn_results}
  \end{center}    
  \end{minipage}
\end{minipage}
\end{figure*}

As data, we generate 10,000
2D points, each represented by a $(x,y)$ coordinate and drawn from one of three clusters, corresponding to the three areas (colors) of expertise. We begin by sampling $\mu \sim $Unif$[0,1]$ and constructing a 2D diagonal matrix $\Sigma$, with diagonal entries sampled from Unif$[0,1]$.
Points are then sampled roughly equally from the three clusters as follows:
$\mathcal{N}(\mu, \Sigma)$ ({\color{orange}orange}), 
$\mathcal{N}(\mu+2.5, \Sigma)$ ({\color{blue}blue}), and 
$\mathcal{N}(\mu+5, \Sigma)$ ({\color{teal}green}). 
Every point is randomly assigned either label 0 (`$\bullet$') or 1 (`+'). \textbf{Figure~\ref{fig:syn_examples}} shows the dataset.
Because class labels are assigned randomly, a classifier knowing only a point's $(x,y)$ coordinates can only achieve 50\% accuracy.
Similarly, a task allocator knowing only the $(x,y)$ coordinates has a 1/3 chance of assigning the input to the correct expert.

\noindent
\textbf{Specifying $dSim$.} Let $e_c$ denote the expert corresponding to color $c$. For input $x$, the optimal $dSim(e_c, x)$ would be 1 when $x$ has color $c$ and 0 otherwise, perfectly assigning each example to the appropriate expert. To investigate the effect of varying informativeness of prior information, we introduce noise parameter $s\in [0, 2/3]$ and define $dSim(e_c, x)$ as follows:
$dSim(e_c, x)=1-s$ if $x$ has color $c$ and $s/2$ otherwise.
With no noise ($s=0$), we revert to the optimal $dSim(e_c, x)$ specified above.   Maximal noise ($s=2/3$) yields $\forall_x dSim(e_c, x)=1/3$: a uniform distribution over all annotators. 
Additional methodological details are provided in Appendix~\ref{sec:syn_appendix}.

\noindent
\textbf{Baselines.} %
\underline{(1) \citet{goel2019crowdsourcing}} learn a task allocation policy using accuracy estimates for all annotators from a history of gold standard tasks. Because we assume that gold standard tasks are unavailable, we instead run their algorithm with accuracy estimates derived using noisy, aggregated annotator predictions.
\underline{(2) \citet{tran2014efficient}} learn an allocation using a multi-arm bandit approach, with initial exploration steps to estimate annotator accuracies followed by exploitation steps that assign inputs to annotators using estimated accuracies.
Once again, in absence of gold standard, the accuracy estimates in the exploration step of their algorithm are obtained using aggregated predictions from the annotators.
\underline{(3) \citet{keswani2021towards}}'s method is equivalent to training an input-specific task allocation model using Algorithm~\ref{alg:main_2} without any prior information, i.e., without $dSim$. In this case, the training algorithm will start with a random allocation. Lacking prior information, we expect {all three of these} baselines to struggle in the closed-training setting we consider. 
See Appendix~\ref{sec:syn_appendix} for further implementation details.

\noindent
\textbf{Results.} We first assume moderate noise $s=0.3$, roughly the middle of $s\in [0, 2/3]$. \textbf{Table~\ref{tbl:syn_results}}
compares \ref{alg:main} and \ref{alg:main_2} vs.\ 
baseline algorithms \cite{goel2019crowdsourcing,tran2014efficient,keswani2021towards}.
We observe large differences in mean label accuracy: 0.90, 0.79, 0.50, 0.50, and 0.41, respectively. Standard error over 50 trials shows that differences are statistically significant. As noted earlier, our assumption of oracle experts means that task allocation accuracy largely determines label accuracy, as the task allocation results here confirm: 0.87, 0.74, 0.33, 0.33, and 0.17, respectively.
The vast difference in task allocation accuracy in training with $dSim$ (\ref{alg:main_2}, \ref{alg:main}) and without $dSim$ (\citet{keswani2021towards} baseline) shows the critical importance of prior information in training. %
The weaker accuracy of \ref{alg:main} vs.\ \ref{alg:main_2} can be explained by their differing use of $dSim$. Whereas \ref{alg:main_2} exploits $dSim$ for initialization only, \ref{alg:main_2} continues to benefit from $dSim$ by utilizing it throughout training. 

\textbf{Figure~\ref{fig:acc_vs_dSim_syn_1}} in Appendix~\ref{sec:syn_appendix}
shows performance for varying $s$ values. As expected, increasing values for noise $s$ leads to a corresponding decrease in accuracy, and for large values of $s$, $dSim$ degrades toward the uniform distribution as an uninformative prior.

\section{Evaluating Task Allocation on a Real-world Dataset for Toxicity Detection} \label{eval-real}

Civil Comments \cite{borkan2019nuanced} provides toxicity labels for 1.8M news comments. Of these, 450K comments are also labeled for the demographic group targeted (e.g., LGBTQ+, race, etc.). We consider the binary classification task of predicting whether a comment contains an ``identity attack'': a comment that is toxic and targets a specific demographic affiliation. 
\citet{goyal2022your} augment Civil Comments with additional demographic identity labels of the annotators.
This enables study of how annotator identity may influence their toxicity ratings.
They sample 25.5k comments to augment with additional labels, uniformly sampling comments from three targeted groups: LGBTQ, African-American, and Control (identity agnostic). Of these, 12\% of comments are labelled as containing an identity attack. Roughly 1K crowd annotators contributed labels to their study, with around 1/3 of annotators affiliated with each demographic group. Each comment is labeled by 5 annotators from each group (i.e., 15  in total). See Appendix~\ref{sec:real_appendix} for additional details.

\begin{table}[t]
    \centering
    \begin{tabular}{l|l}
    \toprule
         Identity mentioned &  Post\\
         \midrule
         LGBTQ & \lb{``I am NOT protecting or condoning the alleged\\ behavior! I’m pointing out the ‘he did\\ this because he is gay' bigotry.''\\}\\
         LGBTQ & ``I feel the same fear for the gay members of my family.'' \\
         \midrule
         African-American & \lb{``I'm sure it was merely an oversight but...not\\ mentioned in the story is that the killer was\\ black and the victims were white. Jus' sayin'.''\\}\\
         African-American & \lb{``You apparently can say whatever you want about Mexicans, Hispanics \&\\ Black people, but the Republican Party draws the line on white women.''}\\
         \bottomrule
    \end{tabular}
    \caption{Example posts targeting LGBTQ or Black identities in toxicity data. For these posts, there is disagreement in toxicity labels of annotators with different demographics \cite{goyal2022your}.
    } 
    \label{tbl:examples}
\end{table}

Control annotators often label toxicity differently than annotators whose own demographic group is targeted in a comment. \textbf{Table~\ref{tbl:examples}} shows illustrative examples. Such differences in toxicity ratings by annotator demographic indicate a form of consistent bias, motivating our consideration of demographics in task allocation.

\noindent
\textbf{Specifying $dSim$.} We investigate potential allocation accuracy improvement by matching annotator demographics to the target groups. For  comment $x$ that targets demographic group $g$ and annotator $e$, we define 
 $dSim(z,e) {=} 1$, if $e$ identifies with $g$ and 0 {otherwise}.

\noindent\textbf{Baselines.} %
We evaluate against baseline training algorithms from \citet{goel2019crowdsourcing, tran2014efficient} and \citet{keswani2021towards}.
The descriptions of these baselines are provided in \S\ref{eval-syn-ta}.
See Appendix~\ref{sec:real_appendix} for model and implementation details of our algorithms and the baselines.

\noindent\textbf{Measurement.} 
We follow \citet{goyal2022your} in reporting AUC score: the area under the receiver operating characteristic (ROC) curve. The dataset is skewed (only 12\% of comments contain identity attacks), and AUC  is appropriately sensitive to such class-imbalance. We randomly split the dataset into train-test partitions (70-30 split), evaluate methods across 25 trials of different splits, and report AUC mean and standard error.

\noindent\textbf{Alternative Gold Standards}. We measure label accuracy using two views of ground-truth: 1) the classic assumption of a single, objective gold standard vs.\ 2) that the gold standard is subjective and varies by community \cite{kumar2021designing,gordon2022jury,goyal2022your}. For the objective gold setting, we induce gold by majority vote over all annotators from Civil Comments \cite{borkan2019nuanced}. Note that these annotators used to define the objective gold are completely disjoint from the set of annotators available for our task allocation experiments. In the subjective setting, we induce gold using the majority vote over the 5 annotators in \cite{goyal2022your} whose demographic identity matches the comment's target demographic. Here, the annotators available for task allocation includes the experts whose majority vote determines gold. Again, gold labels are never 
used for training allocation models, but for evaluation purposes only.

\begin{table}[t]
    \centering
    \begin{tabular}{lcc}
        \toprule
        & Objective gold & Subjective gold \\
        Method & AUC Score & AUC Score \\
        \midrule        
        \ref{alg:main_2} & .62 (.01) & .71 (.01)  \\
        \ref{alg:main} & .59 (0) & .67 (.01)\\
        \midrule
        \citet{goel2019crowdsourcing} & .60 (.01)  & .64 (.02) \\
        \citet{tran2014efficient} & .60 (0) & .66 (.01) \\
        \citet{keswani2021towards} & .60 (.01) & .66 (.01) \\
        \bottomrule
    \end{tabular}
    \captionof{table}{{Objective and subjective gold results. We report AUC over 25 trials (standard error in brackets).}}
    \label{tbl:results_real}
    \end{table}
\noindent{\bf Results.} \textbf{Table~\ref{tbl:results_real}} present results for objective and subjective gold conditions. In both settings, training without prior information yields lower accuracy. 
We observe improvement in performance when using prior information despite the fact that differences in annotator accuracies across demographics are not statistically-significant.
This is because, after training, allocation weights for each input contain information from both prior and observed samples, and correspondingly every test input is assigned to the top-ranked annotator for that input.
The accuracy of the top-ranked annotator is often better than the average annotator accuracy, leading to improved prediction scores.

\noindent\textbf{Results: objective gold.} 
\label{sec:results_predefined}
We observe negligible standard error ($\sim0.01$) in AUC scores across trials, indicating consistency of the mean AUC scores for comparing methods. \ref{alg:main_2} achieves the best AUC score (0.62), 2\% better than prior work baselines that lack prior information (i.e., training without $dSim$). \ref{alg:main} performs 1\% worse than these baselines, likely due to insufficiency of using $dSim$ only for initialization. In contrast, \ref{alg:main_2} mitigates this issue by using $dSim$ throughout the training.

\noindent\textbf{Results: subjective gold.} 
\ref{alg:main_2} again achieves the top mean AUC score (0.71), with 5-7\% improvement over baselines. In contrast with the objective gold setting, \ref{alg:main} also outperforms all baselines (1-3\%). In general, we observe both larger margins and higher overall scores than in the objective gold setting. In part, this may reflect a simple dataset artifact
(e.g., all methods perform better in the subjective vs.\ objective gold setting). 
We also noted a minor artifact earlier in experimental design that could inflate scores here: whereas the objective gold setting uses disjoint annotator pools to define gold vs.\ task allocation, here the annotator pool for task allocation also includes the 5 annotators who define the community gold standard. Despite these confounds, strong intuition remains to expect greater benefit from task allocation in the community-gold setting: when community members are empowered to define gold for their community, we stand to benefit more from engaging them as annotators for their community.

\section{Discussion, Limitations, and Future Work} \label{sec:discussion}

    \textbf{Availability of \textit{gold standard}.} As mentioned in \S\ref{sec:training_overview}, prior studies on task allocation often assume that annotator correctness can be accurately determined.
    However, in real-life we only have access to fallible human decisions (no outside oracle exists) and task allocation seeks to find the suitable annotators whose aggregated decision is considered the gold \cite{ipeirotis2010quality}.
    Given that humans define the gold, the assumption that we can accurately determine annotator correctness would not always be true.
    That is not to say correctness can never be determined; for tasks with objectively correct answers, e.g. question-answer tasks \cite{fan2015icrowd}, annotator qualities can be easily measured.
    But when the ground truth is subjective (e.g., in toxicity analysis) or for settings that are yet unexplored through crowdsourcing (e.g., regional language moderation), assuming presence of gold labels can be unrealistic.
    
    The subjectivity of ground truth is also an important factor.
    The predominant view in crowdsourcing, that the \textit{gold standard} is the majority decision of a random group of annotators, has been challenged by many studies \cite{goyal2022your,kumar2021designing,sap2021annotators,sen2015turkers}.
    Community-defined gold standard definitions have thus been forwarded as a way to incorporate minority voices in machine learning \cite{gordon2022jury,sen2015turkers}.
    Our framework, hence, takes a contextual approach to task allocation. 
    Providing annotator demographics and background as prior information ensures that the social context of the tasks is taken into account.
    Training our closed-loop framework does not require any ground truth, ensuring separation from predefined ideas of ``correctness''.
    Finally, evaluation can be performed for both community-based and majority-based gold standards (as in \S~\ref{eval-real}), depending on the application in question.
    
    \noindent
    \textbf{Prior information and $dSim$.}
    Through empirical evaluations, we show that prior information can improve closed-loop model training.
    However, certain settings may not require such prior information for accurate training, e.g. tasks where the ground truth is considered objective (e.g., factual question-answer datasets) or when annotator qualities are not input-specific
    Secondly, providing incorrect or non-contextual prior information to the framework can have a negative impact on the training process.
    Our algorithms assume that $dSim$ similarity is a weak proxy of annotator quality for a given task, and an incorrect estimate of $dSim$ can lead to incorrect allocation in the initial iterations, thus derailing the entire training process (as observed for noisy $dSim$ in \S~\ref{eval-syn-ta}).
    Finally, a key assumption we make is that annotator demographics and target demographics are known. While it is reasonable to expect annotator demographics to be provided (e.g., using in-take surveys during onboarding), demographics associated with the tasks (e.g., groups targeted in social media posts) may not always be available. In practice, target demographic would have to be manually labeled or automatically detected (with noise). 
    Tackling noise in target demographics merits future exploration and can improve  our framework's applicability.

    \noindent
    \textbf{Annotator consultation costs.} 
    Different annotators can have different consultation costs.
    E.g., platforms like Upwork
    \cite{green2018fueling} allow clients to employ the human experts (or freelancers) for their posted jobs. Experts often have more experience/training (and higher prices) than generalist workers.
    Our framework supports each person having such additional input-specific costs.
    Let  $c_{e_j}: \mathcal{X} \rightarrow \R$ for $e_j$ denote the input-specific cost function for annotator $e_j$.
    Recall from \S\ref{sec:training_overview} that the loss function for the task allocation model is captured by $\mathcal{L}_D$.
    To incorporate input-specific costs, we can alternately minimize a regularized loss function: $\mathcal{L} := \mathcal{L}_D + \lambda \cdot \E_{x}\left[D(x)^\top c(x)\right]$,
    where $\lambda \geq 0$ is the cost hyperparameter.
    Minimizing this cost-regularized loss will ensure that the annotator costs are accounted.

    \noindent
    \textbf{Updating annotators.}
    To add a new annotator, we can assign them weight proportional to their $dSim$ value for any given input category and subsequent training will update this weight 
    based on their predictions.
    Removing an annotator, however, can affect performance if this annotator had expertise in subspaces where all other annotators are inaccurate.
    If the removed annotator did not have any unique expertise, 
    then choosing a large committee size can partially ameliorate this issue.
    However, alternate methods to deal with annotator removal would be beneficial in practice and merit %
    further exploration.

\section{Related Work} \label{sec:related_work}

    Multiple studies in crowdsourcing have evaluated the importance of repeated labelling and proposed methods to handle annotator heterogeneity \cite{sheng2008get,ipeirotis2010quality,liu2012variational,zhou2012learning,welinder2010multidimensional, karger2014budget, bonald2017minimax, goel2019crowdsourcing}.
    Traditionally, most studies consider heterogeneity amongst annotators but not amongst the input tasks.
    In contrast, our framework constructs input-specific allocation models.
    Certain recent papers propose methods to handle correlated heterogeneity of annotators and tasks 
    in offline settings, where all annotators provide a decision for most tasks and the goal is to infer ground truth from given annotations \cite{zhang2019multiple,tao2020label,davani2022dealing}.
    However, offline settings can be expensive as it requires a large number of annotations.
    We tackle online settings where the relevant experts are chosen judiciously to obtain relatively larger cost-benefit.
    
    Studies on online allocation forward a variety of methods to construct appropriate allocation models.
    \citet{yin2020matchmaker} design budget-limited allocation policies that match the annotator preferences to task requirements.
    However, annotator preferences can be unavailable and would be susceptible to implicit biases. 
\citet{liu2012cdas,li2014wisdom} propose frameworks that learn annotator accuracies by comparing to ground truth for completed tasks.
\S\ref{sec:training_overview} shows that such frameworks perform poorly in closed-loop settings when accuracies are estimated using noisy human labels; they can potentially be employed if our proposed approach of using prior information is incorporated into their designs.
\citet{fan2015icrowd} develop allocation models that assigns a task to those annotators who have past experience with similar tasks.
However, estimating inter-task similarity in real-world settings can be expensive due to the large size and variability across tasks.
Certain studies estimate annotator cognitive abilities or use their social network profiles to allocate tasks appropriately \cite{hettiachchi2020crowdcog, goncalves2017task, difallah2013pick}. 
In absence of subsequent training, these methods will have low accuracy when the annotator profiles have insufficient information about their qualities.
Our framework, instead, employs simpler demographic and task-related information to pre-assess worker qualities and complements it with subsequent training.
    \citet{ho2013adaptive} and \citet{ho2012online} study a different setup where human annotators arrive in an online fashion and provide allocation approaches when annotators' skill levels are known in advance.
    Bandit approaches  \cite{arora2012multiplicative,lu2010contextual,valera2018enhancing,tran2014efficient} can also be used for task allocation but primarily assume access to true rewards/penalties. In the case of closed-loop models, they suffer from exacerbated inaccuracies in heterogeneous settings, as observed in \S\ref{eval-syn-ta}.
    
    Recent  human-in-the-loop research studies deferral frameworks that train an automated classifier which can either make a prediction or defer the decision to a human \cite{keswani2021towards,mozannar2020consistent,madras2018learning}.
    Deciding whether a prediction should be made by the classifier or (one or more) humans is a task allocation problem.
    However, here again
    prior algorithms for deferral training assume access to true class labels for training \cite{keswani2021towards,mozannar2020consistent,madras2018learning,hemmer2022forming}, which are unavailable in our closed-loop setting.
    In case of limited ground truth information, semi-supervised classification \cite{chapelle2010semi, triguero2015self, nguyen2015combining, yan2011active, kajino2012convex} selectively use either the model's prediction or labels from noisy crowd annotators to appropriately re-train the model. 
    While the goal of these approaches is to train a classifier, our primary goal is to train an input-specific task allocation model that learns every human decision-maker's error region.
    
\section{Conclusion} \label{sec:conclusion}

We initiate a study of a closed-loop online task allocation framework where decisions from the human annotators are used to continuously train the allocation model as well.
We provide algorithms that utilize the available prior information about the annotators to bootstrap an accurate training process.
By encoding prior information about the human annotators, e.g. demographics and background, we ensure that the learned allocation models are contextually-relevant.
\footnote{Code available at the following link: \url{https://github.com/vijaykeswani/Closed-Loop-Task-Allocation}.}

\clearpage

\bibliographystyle{vancouver-authoryear}
\bibliography{references}

\begin{thebibliography}{59}
\providecommand{\natexlab}[1]{#1}
\providecommand{\url}[1]{\texttt{#1}}
\providecommand{\urlprefix}{}

\bibitem[{Anagnostopoulos et~al.(2012)Anagnostopoulos, Aris and Becchetti, Luca
  and Castillo, Carlos and Gionis, Aristides and Leonardi,
  Stefano}]{anagnostopoulos2012online}
Anagnostopoulos A, Becchetti L, Castillo C, Gionis A, Leonardi S.
\newblock Online team formation in social networks.
\newblock In: Proceedings of the 21st international conference on World Wide
  Web; 2012. p. 839--848.

\bibitem[{Arora et~al.(2012)Arora, Sanjeev and Hazan, Elad and Kale,
  Satyen}]{arora2012multiplicative}
Arora S, Hazan E, Kale S.
\newblock The multiplicative weights update method: A meta-algorithm and
  applications.
\newblock Theory of Computing 2012;8(1):121--164.

\bibitem[{Awerbuch and Kleinberg(2008)Awerbuch, Baruch and Kleinberg,
  Robert}]{awerbuch2008online}
Awerbuch B, Kleinberg R.
\newblock Online linear optimization and adaptive routing.
\newblock Journal of Computer and System Sciences 2008;74(1):97--114.

\bibitem[{Bertrand and Mullainathan(2004)Bertrand, Marianne and Mullainathan,
  Sendhil}]{bertrand2004emily}
Bertrand M, Mullainathan S.
\newblock Are Emily and Greg more employable than Lakisha and Jamal? A field
  experiment on labor market discrimination.
\newblock American economic review 2004;94(4):991--1013.

\bibitem[{Bhuiyan et~al.(2020)Bhuiyan, Md Momen and Zhang, Amy X and Sehat,
  Connie Moon and Mitra, Tanushree}]{bhuiyan2020investigating}
Bhuiyan MM, Zhang AX, Sehat CM, Mitra T.
\newblock Investigating differences in crowdsourced news credibility
  assessment: Raters, tasks, and expert criteria.
\newblock Proceedings of the ACM on Human-Computer Interaction
  2020;4(CSCW2):1--26.

\bibitem[{Bonald and Combes(2017)Bonald, Thomas and Combes,
  Richard}]{bonald2017minimax}
Bonald T, Combes R.
\newblock A minimax optimal algorithm for crowdsourcing.
\newblock Advances in Neural Information Processing Systems 2017;30.

\bibitem[{Borkan et~al.(2019)Borkan, Daniel and Dixon, Lucas and Sorensen,
  Jeffrey and Thain, Nithum and Vasserman, Lucy}]{borkan2019nuanced}
Borkan D, Dixon L, Sorensen J, Thain N, Vasserman L.
\newblock Nuanced metrics for measuring unintended bias with real data for text
  classification.
\newblock In: Companion proceedings of the 2019 world wide web conference;
  2019. p. 491--500.
\newblock
  Dataset:\url{https://www.kaggle.com/competitions/jigsaw-toxic-comment-classification-challenge/data}.

\bibitem[{Chapelle et~al.(2010)Chapelle, Olivier and Sch{\"o}lkopf, Bernhard
  and Zien, Alexander}]{chapelle2010semi}
Chapelle O, Sch{\"o}lkopf B, Zien A.
\newblock Semi-supervised Learning.
\newblock MIT Press; 2010.

\bibitem[{Davani et~al.(2022)Davani, Aida Mostafazadeh and D{\'\i}az, Mark and
  Prabhakaran, Vinodkumar}]{davani2022dealing}
Davani AM, D{\'\i}az M, Prabhakaran V.
\newblock Dealing with disagreements: Looking beyond the majority vote in
  subjective annotations.
\newblock Transactions of the Association for Computational Linguistics
  2022;10:92--110.

\bibitem[{Davidson et~al.(2019)Davidson, Thomas and Bhattacharya, Debasmita and
  Weber, Ingmar}]{davidson2019racial}
Davidson T, Bhattacharya D, Weber I.
\newblock Racial Bias in Hate Speech and Abusive Language Detection Datasets.
\newblock In: Proceedings of the Workshop on Abusive Language Online; 2019. .

\bibitem[{Dawid and Skene(1979)Dawid, Alexander Philip and Skene, Allan
  M}]{dawid1979maximum}
Dawid AP, Skene AM.
\newblock Maximum likelihood estimation of observer error-rates using the EM
  algorithm.
\newblock Journal of the Royal Statistical Society: Series C (Applied
  Statistics) 1979;28(1):20--28.

\bibitem[{Difallah et~al.(2013)Difallah, Djellel Eddine and Demartini, Gianluca
  and Cudr{\'e}-Mauroux, Philippe}]{difallah2013pick}
Difallah DE, Demartini G, Cudr{\'e}-Mauroux P.
\newblock Pick-a-crowd: tell me what you like, and i'll tell you what to do.
\newblock In: Proceedings of the 22nd international conference on World Wide
  Web; 2013. p. 367--374.

\bibitem[{Fan et~al.(2015)Fan, Ju and Li, Guoliang and Ooi, Beng Chin and Tan,
  Kian-lee and Feng, Jianhua}]{fan2015icrowd}
Fan J, Li G, Ooi BC, Tan Kl, Feng J.
\newblock icrowd: An adaptive crowdsourcing framework.
\newblock In: Proceedings of the 2015 ACM SIGMOD international conference on
  management of data; 2015. p. 1015--1030.

\bibitem[{Fontenla-Romero et~al.(2013)Fontenla-Romero, {\'O}scar and
  Guijarro-Berdi{\~n}as, Bertha and Martinez-Rego, David and
  P{\'e}rez-S{\'a}nchez, Beatriz and Peteiro-Barral,
  Diego}]{fontenla2013online}
Fontenla-Romero {\'O}, Guijarro-Berdi{\~n}as B, Martinez-Rego D,
  P{\'e}rez-S{\'a}nchez B, Peteiro-Barral D.
\newblock Online machine learning.
\newblock In: Efficiency and Scalability Methods for Computational Intellect
  IGI Global; 2013.p. 27--54.

\bibitem[{Goel and Faltings(2019)Goel, Naman and Faltings,
  Boi}]{goel2019crowdsourcing}
Goel N, Faltings B.
\newblock Crowdsourcing with fairness, diversity and budget constraints.
\newblock In: Proceedings of the 2019 AAAI/ACM Conference on AI, Ethics, and
  Society; 2019. .

\bibitem[{Goncalves et~al.(2017)Goncalves, Jorge and Feldman, Michael and Hu,
  Subingqian and Kostakos, Vassilis and Bernstein, Abraham}]{goncalves2017task}
Goncalves J, Feldman M, Hu S, Kostakos V, Bernstein A.
\newblock Task routing and assignment in crowdsourcing based on cognitive
  abilities.
\newblock In: Proceedings of the 26th International Conference on World Wide
  Web Companion; 2017. p. 1023--1031.

\bibitem[{Gordon et~al.(2022)Gordon, Mitchell L and Lam, Michelle S and Park,
  Joon Sung and Patel, Kayur and Hancock, Jeff and Hashimoto, Tatsunori and
  Bernstein, Michael S}]{gordon2022jury}
Gordon ML, Lam MS, Park JS, Patel K, Hancock J, Hashimoto T, et~al.
\newblock Jury learning: Integrating dissenting voices into machine learning
  models.
\newblock In: CHI Conference on Human Factors in Computing Systems; 2022. p.
  1--19.

\bibitem[{Gorwa et~al.(2020)Gorwa, Robert and Binns, Reuben and Katzenbach,
  Christian}]{gorwa2020algorithmic}
Gorwa R, Binns R, Katzenbach C.
\newblock Algorithmic content moderation: Technical and political challenges in
  the automation of platform governance.
\newblock Big Data \& Society 2020;7(1):2053951719897945.

\bibitem[{Goyal et~al.(2022)Goyal, Nitesh and Kivlichan, Ian and Rosen, Rachel
  and Vasserman, Lucy}]{goyal2022your}
Goyal N, Kivlichan I, Rosen R, Vasserman L.
\newblock Is Your Toxicity My Toxicity? Exploring the Impact of Rater Identity
  on Toxicity Annotation.
\newblock {The 26th ACM Conference On Computer-Supported Cooperative Work And
  Social Computing (CSCW)}
  2022;Dataset:\url{https://www.kaggle.com/datasets/google/jigsaw-specialized-rater-pools-dataset}.

\bibitem[{Green et~al.(2018)Green, Daryl D and others}]{green2018fueling}
Green DD, et~al.
\newblock Fueling the gig economy: a case study evaluation of Upwork. com.
\newblock Manag Econ Res J 2018;4(2018):3399.

\bibitem[{Hemmer et~al.(2022)Hemmer, Patrick and Schellhammer, Sebastian and
  V{\"o}ssing, Michael and Jakubik, Johannes and Satzger,
  Gerhard}]{hemmer2022forming}
Hemmer P, Schellhammer S, V{\"o}ssing M, Jakubik J, Satzger G.
\newblock Forming Effective Human-AI Teams: Building Machine Learning Models
  that Complement the Capabilities of Multiple Experts.
\newblock arXiv preprint arXiv:220607948 2022;.

\bibitem[{Hettiachchi et~al.(2020)Hettiachchi, Danula and Van Berkel, Niels and
  Kostakos, Vassilis and Goncalves, Jorge}]{hettiachchi2020crowdcog}
Hettiachchi D, Van~Berkel N, Kostakos V, Goncalves J.
\newblock CrowdCog: A Cognitive skill based system for heterogeneous task
  assignment and recommendation in crowdsourcing.
\newblock Proceedings of the ACM on Human-Computer Interaction
  2020;4(CSCW2):1--22.

\bibitem[{Ho et~al.(2013)Ho, Chien-Ju and Jabbari, Shahin and Vaughan, Jennifer
  Wortman}]{ho2013adaptive}
Ho CJ, Jabbari S, Vaughan JW.
\newblock Adaptive Task Assignment for Crowdsourced Classification.
\newblock In: International Conference on Machine Learning; 2013. p. 534--542.

\bibitem[{Ho and Vaughan(2012)Ho, Chien-Ju and Vaughan,
  Jennifer}]{ho2012online}
Ho CJ, Vaughan J.
\newblock Online task assignment in crowdsourcing markets.
\newblock In: Proceedings of the AAAI Conference on Artificial Intelligence,
  vol.~26; 2012. p. 45--51.

\bibitem[{Ipeirotis et~al.(2010)Ipeirotis, Panagiotis G and Provost, Foster and
  Wang, Jing}]{ipeirotis2010quality}
Ipeirotis PG, Provost F, Wang J.
\newblock Quality management on amazon mechanical turk.
\newblock In: Proceedings of the ACM SIGKDD workshop on human computation;
  2010. p. 64--67.

\bibitem[{Kajino et~al.(2012)Kajino, Hiroshi and Tsuboi, Yuta and Kashima,
  Hisashi}]{kajino2012convex}
Kajino H, Tsuboi Y, Kashima H.
\newblock A convex formulation for learning from crowds.
\newblock In: Twenty-Sixth AAAI Conference on Artificial Intelligence; 2012. .

\bibitem[{Karger et~al.(2014)Karger, David R and Oh, Sewoong and Shah,
  Devavrat}]{karger2014budget}
Karger DR, Oh S, Shah D.
\newblock Budget-optimal task allocation for reliable crowdsourcing systems.
\newblock Operations Research 2014;62(1):1--24.

\bibitem[{Keswani et~al.(2021)Keswani, Vijay and Lease, Matthew and Kenthapadi,
  Krishnaram}]{keswani2021towards}
Keswani V, Lease M, Kenthapadi K.
\newblock Towards Unbiased and Accurate Deferral to Multiple Experts.
\newblock In: Proceedings of the AAAI/ACM Conference on AI, Ethics, and
  Society; 2021. .

\bibitem[{Kumar et~al.(2021)Kumar, Deepak and Kelley, Patrick Gage and
  Consolvo, Sunny and Mason, Joshua and Bursztein, Elie and Durumeric, Zakir
  and Thomas, Kurt and Bailey, Michael}]{kumar2021designing}
Kumar D, Kelley PG, Consolvo S, Mason J, Bursztein E, Durumeric Z, et~al.
\newblock Designing toxic content classification for a diversity of
  perspectives.
\newblock In: Seventeenth Symposium on Usable Privacy and Security (SOUPS
  2021); 2021. p. 299--318.

\bibitem[{Li et~al.(2014)Li, Hongwei and Zhao, Bo and Fuxman,
  Ariel}]{li2014wisdom}
Li H, Zhao B, Fuxman A.
\newblock The wisdom of minority: Discovering and targeting the right group of
  workers for crowdsourcing.
\newblock In: Proceedings of the 23rd international conference on World wide
  web; 2014. p. 165--176.

\bibitem[{Liu et~al.(2012{\natexlab{a}})Liu, Qiang and Peng, Jian and Ihler,
  Alexander T}]{liu2012variational}
Liu Q, Peng J, Ihler AT.
\newblock Variational inference for crowdsourcing.
\newblock Advances in neural information processing systems 2012;25.

\bibitem[{Liu et~al.(2012{\natexlab{b}})Liu, Xuan and Lu, Meiyu and Ooi, Beng
  Chin and Shen, Yanyan and Wu, Sai and Zhang, Meihui}]{liu2012cdas}
Liu X, Lu M, Ooi BC, Shen Y, Wu S, Zhang M.
\newblock CDAS: A Crowdsourcing Data Analytics System.
\newblock Proceedings of the VLDB Endowment 2012;5(10).

\bibitem[{Lu et~al.(2010)Lu, Tyler and P{\'a}l, D{\'a}vid and P{\'a}l,
  Martin}]{lu2010contextual}
Lu T, P{\'a}l D, P{\'a}l M.
\newblock Contextual multi-armed bandits.
\newblock In: Proceedings of the Thirteenth international conference on
  Artificial Intelligence and Statistics JMLR Workshop and Conference
  Proceedings; 2010. p. 485--492.

\bibitem[{Madras et~al.(2018)Madras, David and Creager, Elliot and Pitassi,
  Toniann and Zemel, Richard}]{madras2018learning}
Madras D, Creager E, Pitassi T, Zemel R.
\newblock Learning adversarially fair and transferable representations.
\newblock arXiv preprint arXiv:180206309 2018;.

\bibitem[{Mehrabi et~al.(2019)Mehrabi, Ninareh and Morstatter, Fred and Saxena,
  Nripsuta and Lerman, Kristina and Galstyan, Aram}]{mehrabi2019survey}
Mehrabi N, Morstatter F, Saxena N, Lerman K, Galstyan A.
\newblock A survey on bias and fairness in machine learning.
\newblock arXiv preprint arXiv:190809635 2019;.

\bibitem[{Mozannar and Sontag(2020)Mozannar, Hussein and Sontag,
  David}]{mozannar2020consistent}
Mozannar H, Sontag D.
\newblock Consistent estimators for learning to defer to an expert.
\newblock In: International Conference on Machine Learning PMLR; 2020. p.
  7076--7087.

\bibitem[{Nguyen et~al.(2015)Nguyen, An Thanh and Wallace, Byron C and Lease,
  Matthew}]{nguyen2015combining}
Nguyen AT, Wallace BC, Lease M.
\newblock Combining crowd and expert labels using decision theoretic active
  learning.
\newblock In: Third AAAI conference on human computation and crowdsourcing;
  2015. .

\bibitem[{Pan et~al.(2021)Pan, Yuan and Froese, Fabian and Liu, Ni and Hu,
  Yunyang and Ye, Maolin}]{pan2021adoption}
Pan Y, Froese F, Liu N, Hu Y, Ye M.
\newblock The adoption of artificial intelligence in employee recruitment: The
  influence of contextual factors.
\newblock The International Journal of Human Resource Management 2021;p. 1--23.

\bibitem[{Pennington et~al.(2014)Pennington, Jeffrey and Socher, Richard and
  Manning, Christopher D}]{pennington2014glove}
Pennington J, Socher R, Manning CD.
\newblock {GloVe}: Global Vectors for Word Representation.
\newblock In: Proceedings of the 2014 conference on empirical methods in
  natural language processing (EMNLP); 2014. p. 1532--1543.

\bibitem[{Raghu et~al.(2019)Raghu, Maithra and Blumer, Katy and Sayres, Rory
  and Obermeyer, Ziad and Kleinberg, Bobby and Mullainathan, Sendhil and
  Kleinberg, Jon}]{raghu2019direct}
Raghu M, Blumer K, Sayres R, Obermeyer Z, Kleinberg B, Mullainathan S, et~al.
\newblock Direct uncertainty prediction for medical second opinions.
\newblock In: International Conference on Machine Learning; 2019. p.
  5281--5290.

\bibitem[{Sap et~al.(2019)Sap, Maarten and Card, Dallas and Gabriel, Saadia and
  Choi, Yejin and Smith, Noah A}]{sap2019risk}
Sap M, Card D, Gabriel S, Choi Y, Smith NA.
\newblock The risk of racial bias in hate speech detection.
\newblock In: Proceedings of ACL; 2019. p. 1668--1678.

\bibitem[{Sap et~al.(2022)Sap, Maarten and Swayamdipta, Swabha and Vianna,
  Laura and Zhou, Xuhui and Choi, Yejin and Smith, Noah A}]{sap2021annotators}
Sap M, Swayamdipta S, Vianna L, Zhou X, Choi Y, Smith NA.
\newblock Annotators with attitudes: How annotator beliefs and identities bias
  toxic language detection.
\newblock NAACL 2022;.

\bibitem[{Schumann et~al.(2020)Schumann, Candice and Foster, Jeffrey and
  Mattei, Nicholas and Dickerson, John}]{schumann2020we}
Schumann C, Foster J, Mattei N, Dickerson J.
\newblock We need fairness and explainability in algorithmic hiring.
\newblock In: International Conference on Autonomous Agents and Multi-Agent
  Systems (AAMAS); 2020. .

\bibitem[{Sch{\"u}tze et~al.(2008)Sch{\"u}tze, Hinrich and Manning, Christopher
  D and Raghavan, Prabhakar}]{schutze2008introduction}
Sch{\"u}tze H, Manning CD, Raghavan P.
\newblock Introduction to information retrieval, vol.~39.
\newblock Cambridge University Press Cambridge; 2008.

\bibitem[{Sen et~al.(2015)Sen, Shilad and Giesel, Margaret E and Gold, Rebecca
  and Hillmann, Benjamin and Lesicko, Matt and Naden, Samuel and Russell, Jesse
  and Wang, Zixiao and Hecht, Brent}]{sen2015turkers}
Sen S, Giesel ME, Gold R, Hillmann B, Lesicko M, Naden S, et~al.
\newblock Turkers, scholars," arafat" and" peace" cultural communities and
  algorithmic gold standards.
\newblock In: Proceedings of the 18th acm conference on computer supported
  cooperative work \& social computing; 2015. p. 826--838.

\bibitem[{Sheng et~al.(2008)Sheng, Victor S and Provost, Foster and Ipeirotis,
  Panagiotis G}]{sheng2008get}
Sheng VS, Provost F, Ipeirotis PG.
\newblock Get another label? improving data quality and data mining using
  multiple, noisy labelers.
\newblock In: Proceedings of the 14th ACM SIGKDD international conference on
  Knowledge discovery and data mining; 2008. p. 614--622.

\bibitem[{Spinde et~al.(2021)Spinde, Timo and Rudnitckaia, Lada and Sinha,
  Kanishka and Hamborg, Felix and Gipp, Bela and Donnay,
  Karsten}]{spinde2021mbic}
Spinde T, Rudnitckaia L, Sinha K, Hamborg F, Gipp B, Donnay K.
\newblock MBIC--A Media Bias Annotation Dataset Including Annotator
  Characteristics.
\newblock iConference 2021;.

\bibitem[{Surowiecki(2005)Surowiecki, James}]{surowiecki2005wisdom}
Surowiecki J.
\newblock The wisdom of crowds.
\newblock Anchor; 2005.

\bibitem[{Tao et~al.(2020)Tao, Fangna and Jiang, Liangxiao and Li,
  Chaoqun}]{tao2020label}
Tao F, Jiang L, Li C.
\newblock Label similarity-based weighted soft majority voting and pairing for
  crowdsourcing.
\newblock Knowledge and Information Systems 2020;62(7):2521--2538.

\bibitem[{Tran-Thanh et~al.(2014)Tran-Thanh, Long and Stein, Sebastian and
  Rogers, Alex and Jennings, Nicholas R}]{tran2014efficient}
Tran-Thanh L, Stein S, Rogers A, Jennings NR.
\newblock Efficient crowdsourcing of unknown experts using bounded multi-armed
  bandits.
\newblock Artificial Intelligence 2014;214:89--111.

\bibitem[{Triguero et~al.(2015)Triguero, Isaac and Garc{\'\i}a, Salvador and
  Herrera, Francisco}]{triguero2015self}
Triguero I, Garc{\'\i}a S, Herrera F.
\newblock Self-labeled techniques for semi-supervised learning: taxonomy,
  software and empirical study.
\newblock Knowledge and Information systems 2015;42(2):245--284.

\bibitem[{Valcarce et~al.(2016)Valcarce, Daniel and Parapar, Javier and
  Barreiro, {\'A}lvaro}]{valcarce2016additive}
Valcarce D, Parapar J, Barreiro {\'A}.
\newblock Additive smoothing for relevance-based language modelling of
  recommender systems.
\newblock In: Proceedings of the 4th Spanish Conference on Information
  Retrieval; 2016. p. 1--8.

\bibitem[{Valera et~al.(2018)Valera, Isabel and Singla, Adish and Gomez
  Rodriguez, Manuel}]{valera2018enhancing}
Valera I, Singla A, Gomez~Rodriguez M.
\newblock Enhancing the Accuracy and Fairness of Human Decision Making.
\newblock Advances in Neural Information Processing Systems 2018;31:1769--1778.

\bibitem[{Welinder et~al.(2010)Welinder, Peter and Branson, Steve and Perona,
  Pietro and Belongie, Serge}]{welinder2010multidimensional}
Welinder P, Branson S, Perona P, Belongie S.
\newblock The multidimensional wisdom of crowds.
\newblock Advances in neural information processing systems 2010;23.

\bibitem[{Wu et~al.(2021)Wu, Gang and Chen, Zhiyong and Liu, Jia and Han,
  Donghong and Qiao, Baiyou}]{wu2021task}
Wu G, Chen Z, Liu J, Han D, Qiao B.
\newblock Task assignment for social-oriented crowdsourcing.
\newblock Frontiers of Computer Science 2021;15(2):1--11.

\bibitem[{Yan et~al.(2011)Yan, Yan and Rosales, Romer and Fung, Glenn and Dy,
  Jennifer G}]{yan2011active}
Yan Y, Rosales R, Fung G, Dy JG.
\newblock Active learning from crowds.
\newblock In: International Conference of Machine Learning; 2011. .

\bibitem[{Yin et~al.(2020)Yin, Xiaoyan and Chen, Yanjiao and Xu, Cheng and Yu,
  Sijia and Li, Baochun}]{yin2020matchmaker}
Yin X, Chen Y, Xu C, Yu S, Li B.
\newblock Matchmaker: Stable Task Assignment With Bounded Constraints for
  Crowdsourcing Platforms.
\newblock IEEE Internet of Things Journal 2020;8(3).

\bibitem[{Zhang et~al.(2019)Zhang, Hao and Jiang, Liangxiao and Xu,
  Wenqiang}]{zhang2019multiple}
Zhang H, Jiang L, Xu W.
\newblock Multiple Noisy Label Distribution Propagation for Crowdsourcing.
\newblock In: IJCAI; 2019. p. 1473--1479.

\bibitem[{Zhou et~al.(2012)Zhou, Dengyong and Basu, Sumit and Mao, Yi and
  Platt, John}]{zhou2012learning}
Zhou D, Basu S, Mao Y, Platt J.
\newblock Learning from the wisdom of crowds by minimax entropy.
\newblock Advances in neural information processing systems 2012;25.

\end{thebibliography}

\clearpage
\appendix

\section{Proofs} \label{sec:proofs}
For our two algorithms using the $dSim$ prior, we show that the final trained allocation model converges to a point that simulates the underlying accuracy functions of the annotators.
This holds even if we start with a ``weak'' $dSim$ function.
In particular, if any annotator $e_j$ has high accuracy for category $z$, then how fast the algorithm converges to an allocation model that assigns high weight to $e_j$ for category $z$ depends on the initial weight assigned to $e_j$ for $z$.

\begin{theorem}[Exploitation using $dSim$] \label{thm:exploitation}
For group $z$, assume annotator $e_j$ is more accurate than all others. %
For $\beta > 0$, suppose we set $dSim$ function {\em st.} $dSim(e_j, z) - \max_{j' \in \set{1, \dots, m} \setminus \set{j}} dSim(e_{j'}, z) \geq \beta$.
Assume all annotators receive the same rewards or penalties for correct/incorrect predictions. 
Then the training algorithm that initializes $D(x)$ parameters $u$ with this $dSim$ function increases the weight assigned to annotator $e_j$ by at least $2\beta \delta$ in expectation, where $\delta \in [0,1]$ depends on the choice of $\der$ and $\dep$ values.
\end{theorem}

\textit{Proof.} For input group $z$, let $d_i := dSim(e_i, z)$. 
Since rewards and penalties are the same for all annotators for this input, we denote the reward by $\der$ and penalty by $\dep$ for simplicity.
By the condition in the theorem, for $j' \in \set{1, \dots, m} \setminus \set{j}$, we have that $d_{j'} < d_j - \beta$.
Then, if we select a single annotator for prediction, the expected change in annotator $j$'s weight is at least
$d_j \der^{(j)} - \left( \sum_{j' \neq j} (d_{j'} - \beta) \right) \dep.$
The first term is the reward if annotator $j$ is selected for making the decision while the second term is the penalty if some other expert is selected.
Since the overall updated weights have to be normalized, $\delta := \der = (m-1) \dep$.
Therefore,
the expected change annotator $j$'s weight is at least
\begin{align*}
d_j \delta - \left( \sum_{j' \neq j} (d_{j'} - \beta) \right) &\frac{\delta}{m-1}
= d_j \delta - \frac{\delta}{m-1} \sum_{j' \neq j} d_{j'} + \delta \beta \\
&= \frac{\delta}{m-1} \sum_{j' \neq j} (d_j - d_{j'}) + \delta \beta \geq 2\delta \beta.
\end{align*}    
\noindent
Next, we show that, for any input category $z$, even if there are accurate annotators who are not assigned high weight by $dSim$ function, they can be ``discovered'' by the training algorithm.

\begin{theorem}[Exploration of accurate annotators] \label{thm:exploration}
For any input group $z$, assume annotator $e_j$ has perfect accuracy ($1$).
Let $k$ be the size of the committee sampled from $D(x)$ to label input $x$.
Let the $dSim$ function be set such that $dSim(e_j, z) = \epsilon$, for some  $\epsilon \in [0,1]$, but the total weight (normalized) assigned by $dSim$ to accurate annotators for group $z$ is greater than 0.5. 
Assume all annotators receive the same rewards for correct prediction and same penalties for incorrect prediction. 
Then, there is an expected positive increase in the weight of this annotator if
$\epsilon > 1 - \left(1 - \frac{k}{2m} \right)^{1/k}$.
\end{theorem}
\textit{Proof.} As rewards and penalties are the same for all annotators for this input, we denote the reward by $\der$ and penalty by $\dep$ for simplicity.
Probability that annotator $e_j$ is not selected in each of the $k$ samples is $(1-\epsilon)^k$. Since the total weight assigned by $dSim$ to accurate annotators for group $z$ is greater than 0.5, this annotator will be rewarded only if they are selected and otherwise penalized (not being selected leads to weight penalization since the weights are normalized to form a distribution).
Therefore, the expected change in the weight of annotator $e_j$ is
$(1-(1-\epsilon)^k) \der - (1-\epsilon)^k \dep.$
On expectation, majority of annotators in the chosen committee are correct, hence, due to the normalization constraint, $\der = (m/k'-1)\dep$, where $k' > k/2$ are the number of correct annotators.
Let $\epsilon' := (1-(1-\epsilon)^k)$.
For the expected change to be positive, we need $epsilon' \der > (1-\epsilon') \dep$
\begin{align*}
\implies & \epsilon' \left(\frac{2m}{k} -1 \right) \dep > (1-\epsilon') \dep \\
\implies &  \frac{2m}{k} -1 > \frac{1}{\epsilon'}-1 \implies \epsilon' > \frac{k}{2m}
\implies (1-(1-\epsilon)^k) > \frac{k}{2m}\\ \implies &  1-\epsilon < \left(1 - \frac{k}{2m} \right)^{1/k} 
\implies \epsilon > 1 - \left(1 - \frac{k}{2m} \right)^{1/k}.
\end{align*}

\noindent
\textbf{Proof of Claim~\ref{clm:example}.}
Recall that $\alpha > 0.5$ fraction of annotators are biased against group $z=0$ and $(1-\alpha)$ fraction are biased against group $z=1$.
Assuming no prior, the training starts with an allocation that assigns uniform weight $1/m$ to all annotators. 
When $k=1$, the allocation model chooses a single expert to make the final decision.
Starting accuracy for group $z=1$ elements is $\alpha + 0.5(1-\alpha)$ and the starting accuracy for group $z=0$ elements is $(1-\alpha) + 0.5\alpha $. Hence the disparity between the two groups at step 0 is $(\alpha-0.5)$.
In the first training step, suppose we see a sample from group $z=0$ (will consider the other case later).
Training using prediction on this sample will have the following impact.
With probability $(1-\alpha)$ we will choose an unbiased annotator and with probability $\alpha$ we will choose a biased annotator.
Since the annotators only provide a binary label (and not a score), we can assume that $\der^{(i)}, \dep^{(i)}$ for correct prediction are the same for all annotators. 
We will denote them by simply $\der,\dep$.
The weight of the chosen annotator is increased by quantity $\der$ and weight of other experts is decreased by quantity $\dep$; for appropriate normalization, $\delta := \der = (m-1)\dep$.
If an unbiased annotator is chosen, the resulting accuracy for group $z=0$ (after the weights of all annotators are updated) is 
\begin{align*}
    &\frac{\alpha m}{2} \left( \frac{1}{m} - \frac{\delta}{m-1} \right) + ((1-\alpha)m - 1) \left( \frac{1}{m} - \frac{\delta}{m-1} \right) + \frac{1}{m} + \delta\\
    =& (1 - 0.5\alpha) + \frac{\delta\alpha m}{2(m-1)} .
\end{align*}
Since an unbiased annotator is chosen, accuracy after one step of training increases by some amount.
If a biased annotator is chosen instead, the resulting accuracy for group $z=0$ is 
\begin{align*}
    &(1-\alpha)m \left( \frac{1}{m} - \frac{\delta}{m-1} \right) + \frac{1}{2}\left( (\alpha m-1)\left( \frac{1}{m} - \frac{\delta}{m-1} \right) + \frac{1}{m} + \delta \right)\\
    =& (1 - 0.5\alpha) - \frac{\delta}{2(m-1)} (1-\alpha)m.
\end{align*}
On expectation, the accuracy for group $z=0$ elements after one step of training (using a $z=0$ sample) is
\[(1 - 0.5\alpha) + (1-\alpha) \frac{\delta\alpha m}{2(m-1)} - \alpha \frac{\delta}{2(m-1)} (1-\alpha)m = (1-0.5\alpha).\]
Since $z=1$ accuracy will remain unchanged in this case, in expectation, the disparity between the accuracies for the two groups remains unchanged despite training.
Note that we did not use the fact that $\alpha > 0.5$. Hence the analysis for the case when we see a $z=1$ element is symmetric.

\noindent
\textbf{Proof of Claim~\ref{rem:example_dsim}.}
This time we start with a non-random allocation, i.e., an allocation induced by an appropriate $dSim$ function (i.e., for input $(x,z)$, we have that allocation output $D(x)_i \propto dSim(e_i, z)$). 
Again we choose one annotator to whom the decision is deferred.
Then,  for any input $x$ from group $z = 0$, the probability that we obtain the true label is 
\[\frac{(1-\alpha) 1}{(1-\alpha) 1 + \alpha \gamma} + \frac{ \alpha \gamma}{(1-\alpha) 1 + \alpha \gamma} \cdot \frac{1}{2} = 1 - \frac{1}{2}\frac{\alpha \gamma}{(1-\alpha) 1 + \alpha \gamma}.\]
Similarly, for any input $x$ from group $z = 1$, the probability that we obtain the true label based on allocation output is 
\[ 1 - \frac{1}{2}\frac{(1-\alpha) \gamma}{(1-\alpha) \gamma + \alpha 1}.\]
Hence, the starting disparity in accuracy for two groups is 
$$\dc := \frac{1}{2}\frac{\alpha \gamma}{(1-\alpha) 1 + \alpha \gamma} - \frac{1}{2}\frac{(1-\alpha) \gamma}{(1-\alpha) \gamma + \alpha 1}.$$
We can derive a lower bound on this quantity as follows
$$\dc \geq  \frac{1}{2}\frac{\alpha \gamma}{(1-\alpha)  + \alpha \gamma} + \frac{1}{2}\frac{(1-\alpha) \gamma}{(1-\alpha)  + \alpha \gamma}   = \frac{1}{2} \frac{\gamma}{1-\alpha + \alpha \gamma} \geq \frac{\gamma}{2}.$$
Similarly, we can also derive an upper bound as follows,
$$\dc \leq \frac{1}{2}\frac{\alpha \gamma}{(1-\alpha)  + \alpha \gamma}  \leq \frac{1}{2}\frac{\alpha \gamma}{(1-\alpha)}.$$
Note that $\alpha/(1-\alpha) > 1$ since $\alpha > 0.5$.
Hence,
\[\frac{\gamma}{2} \leq \dc \leq \frac{1}{2}\frac{\gamma}{(1-\alpha)}.\]

\section{Prior training algorithms} \label{sec:prior_training}

\noindent
\textbf{Multiplicative weights update (MWU) algorithm.} This algorithm estimates a weight distribution over the available annotators based on their predictions.
In particular, weighted majority MWU \cite{arora2012multiplicative} works in an online manner and follows a similar training approach as Eqn~\ref{updates-True}. At iteration $t$, suppose $(w_{e_1}^{(t)}, \cdots, w_{e_m}^{(t)})$ are the weights assigned to the $m$ annotators. For a predefined $\eta \in [0,0.5]$, if annotator $j$ predicts incorrectly in iteration $t$, then its weight is decreased by a factor of $(1-\eta)$; i.e., $w_{e_j}^{(t+1)} = (1-\eta) w_{e_j}^{(t)}$. 
After normalization, this amounts to rewarding the weights of correct annotators and penalizing the weights of incorrect annotators.
An important difference between the setting tackled by MWU and the task allocation setting we consider is the
the necessity of constructing an input-specific task-allocation policy in our setting.

\noindent
\textbf{Multi-arm bandits (MAB).}
Multi-arm bandit approaches can also be used to train allocation models.
These approaches also involve a exploration phase, where accurate annotators are discovered, and an exploitation phase, that uses the learn accuracies of annotators for correct task allocation \cite{tran2014efficient}.
Treating available task information as ``context'' (default features in our setting), one can also apply a contextual multi-arm bandit (CMAB) framework to find context-specific actions to maximize the total reward \cite{lu2010contextual}. %
Considering MAB algorithms observe rewards/penalties for human decision-makers who are selected and provide predictions, this  amounts to making updates in a manner similar to Eqn~\ref{updates-True}. 
In contrast to a CMAB approach, we adopt a supervised learning approach whereas CMAB %
 assume that the payoff is revealed only after an action.

\section{Additional synthetic data details} \label{sec:syn_appendix}

\begin{figure}
    \centering
    \includegraphics[width=0.7\linewidth]{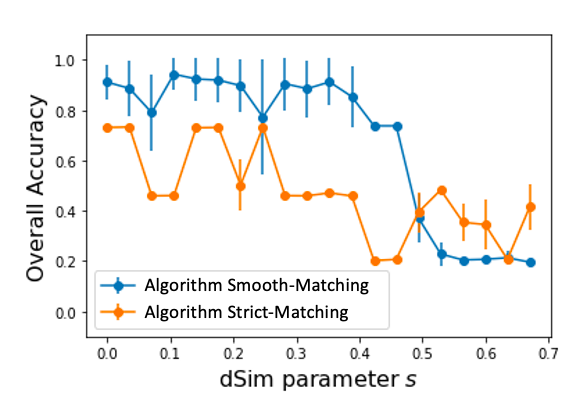}
    \caption{Accuracy vs noise parameter $s$ used in $dSim$ for task allocation experiments presented in \S~\ref{eval-syn-ta}. 
    }
    \label{fig:acc_vs_dSim_syn_1}
\end{figure}

\noindent
\textbf{Allocation model.} The allocation model consists of $m$ individual logistic models.
For each annotator $e$, we construct a distinct logistic model $d_e(x) = \sigma(w_e^\top x)$ for parameters $w_e$. %
For input $x$, the task allocation output is $D(x) = \{d_{e_1}(x), d_{e_2}(x), \ldots, d_{e_m}(x)\}$.
Suppose $C(x)$ is the $k$-sized committee chosen for any input $x$ and suppose $\hy$ is their aggregated decision.
The loss function we use for optimization $\mathcal{L}_D(u) := \E_{x}[ 1/k \sum_{j \in C} (- \mathbf{1}[\hy=e_j(x)] \cdot \log(d_{u,j}(x)) $ $ - \mathbf{1}[\hy\neq e_j(x)]\cdot  \log (1 - d_{u,j}(x))) )].$

\noindent
\textbf{Implementation details for our algorithms.} 
The model update step in both \ref{alg:main} and \ref{alg:main_2} is implemented using using stochastic gradient descent with batch size $B=$ 10. The learning rate $\eta = 0.001$ for \ref{alg:main}.
For \ref{alg:main_2}, learning rate $\eta = 0.1$ and $T_d = 10^4$.
For the implementation of \ref{alg:main}, we need to pre-train the allocation model $D$ so that, for input $x$ from category $z$ and annotator $e$, $D(x)_e \propto dSim(e, z)$.
We use regression over the first 500 unlabeled training samples to perform this pre-training.
That is, for the first $N=1000$ training samples $\set{x_j}_{j=1}^{N}$, we can compute $\set{dSim(e_i, x_j)}_{j=1}^{N}$ for annotator $e_i$. Then we train annotator's model $d_i$ using samples $\set{(x_j, dSim(e_i, x_j))}_{j=1}^{N}$
using stochastic gradient descent with mean-squared loss, learning rate 0.5, and for 1000 iterations.

\noindent
\textbf{Implementation details for baselines.} 
\textit{\citet{goel2019crowdsourcing}.} The approach of \cite{goel2019crowdsourcing} learns a single allocation policy using accuracy estimates of the annotators.
Let $\mathcal{A}: \set{e_1, e_2, \dots, e_m} \times \set{0,1} \rightarrow [0,1]$ denote the function that captures the accuracy estimate of annotator $i$ for every label $y$.
After observing batch of $B$ samples, we have aggregated annotator predictions for these samples - $\set{\hat{y}_j}_{j=1}^B$.
By comparing $e_i$'s predictions over this batch, we can create an empirical estimate for $\mathcal{A}(e_i, 0)$ and $\mathcal{A}(e_i, 1)$, for all $i \in [m]$.
With these estimates, \citet{goel2019crowdsourcing}'s algorithm solves a linear program to update the allocation model.
We solve this linear program using Python's SLSQP library with parameter $\beta = 0.1$.

The approach of \citet{tran2014efficient} uses multi-arm bandit algorithms to efficiently learn an allocation model.
The exploration step basically measures average accuracy of all annotators during training phase (computed with respect to aggregated annotator predictions in our closed-loop setting).
For the test partition, \citet{tran2014efficient} use a greedy algorithm to solve a bounded-knapsack problem for learning the allocation model.
To ensure all inputs are not assigned to one annotator, they impose a budget of $b$ on number of samples each annotator labels. We set this parameter $b = 1000$ (i.e., \#test-samples/\#experts).
For each test sample, the greedy algorithm selects the annotator with highest average accuracy estimate.
If this annotator has already labeled $b$ samples, we move on to the annotator with the next highest average accuracy estimate, and so on.
\textit{\citet{keswani2021towards}} baseline is similar to our algorithms but does not use prior information or $dSim$. Hence, to implement this baseline, we simply run \ref{alg:main} without $dSim$.
All other parameter and implementation details for this baseline are kept to be the same as \ref{alg:main}.

\noindent
\textbf{Other results.} We also study \ref{alg:main} and \ref{alg:main_2} with varying noise parameter $s$ in Figure~\ref{fig:acc_vs_dSim_syn_1}.
As expected, the accuracy of both algorithms decreases with increasing $s$.

\section{Additional toxicity data details} \label{sec:real_appendix}

\noindent
\textbf{Dataset details.} 
The dataset contains 7,036 comments that mention LGBTQ identity 6,341 comments mention African-American identity, and rest are identity-agnostic.
A comment is considered as mentioning LGBTQ identity if it contains words related to the following sexual/gender identities - gay, lesbian, bisexual, transsexual, other gender, other sexual orientation.
Diverse annotators participated (318 LGBTQ, 313 African-American, and 322 Control), with each comment rated by 5 annotators from each group.

\noindent
\textbf{Implementation details for our algorithms.} 
Each input comment $x$ is represented by a vector based on pre-trained, 25D GloVe embeddings \cite{pennington2014glove}. %
For each annotator $e$, we construct a logistic model $d_e(x) = \sigma(w_e^\top x)$ for parameters $w_e$. %
The allocation model $D$ is the set of all of the individual annotator models, i.e., $D(x) = \{d_{e_1}(x), d_{e_2}(x), \ldots, d_{e_m}(x)\}$.
We use a separate model for each annotator to handle sparsity. Because every annotator annotates only a small selection of posts, we need to update a given annotator model only after they provide a label. We use the same loss function $\mathcal{L}_D$ as the synthetic setup and train using stochastic gradient descent with learning rate 0.25 and batch size 1,000.

Using \citet{goyal2022your}'s dataset, we simulate a real-world use-case in which tasks can only be assigned to available annotators. Given an input comment $x$, we compute the task allocation score $D(x)_i$ for each individual $i$. Next, we rank annotators by descending scores and one-by-one (in rank order), request a label from the given annotator. If the annotator is available (i.e., they labeled $x$ in the dataset), we then retrieve their label to use; otherwise we proceed to request a label from the next-ranked annotator. For training, we collect 7 annotations in this manner for each comment, using majority vote as the committee's decision (\S\ref{sec:model}). At test time, however, we request only a single label from top-ranked, available annotator.

The model update step in both \ref{alg:main} and \ref{alg:main_2} is implemented using using stochastic gradient descent with learning rate $\eta =$ 0.25 and batch size $B =$ 1000.
For \ref{alg:main_2}, we set $T_d = 10^6$.
For the implementation of \ref{alg:main}, we again pre-train the allocation model $D$ so that, for input $x$ from category $z$ and annotator $e$, $D(x)_e \propto dSim(e, z)$.
We use regression over the first 500 unlabeled training samples to perform this pre-training, using stochastic gradient descent with mean-squared loss, learning rate 0.5, and for 500 iterations.

\noindent
\textbf{Implementation details for baselines.} 
The implementation of baselines is the same as the one reported in \S\ref{sec:syn_appendix}.
For \citet{goel2019crowdsourcing} baseline, parameter $\beta$ is set to 0.1.
For \citet{tran2014efficient} baseline, budget $b$ is set to be 25.
\citet{keswani2021towards} baseline is implemented using \ref{alg:main_2} with no $dSim$ (i.e., $\mu = 0$).

\end{document}